\documentclass[pre,showpacs,amsfonts,amssymb,eqsecnum,%
twocolumn,floatfix,,superscriptaddress]{revtex4} 
\usepackage{amsmath}
\usepackage{graphicx}
\usepackage{bm}
\allowdisplaybreaks

\begin{document}

\title{Mesoscopic description of random walks on combs}
\author{Vicen\c{c} M\'{e}ndez}
\affiliation{Grup de F{\'i}sica Estad\'{i}stica.  
Departament de F{\'i}sica. Facultat de Ci{\`e}ncies. 
Edifici Cc. Universitat Aut\`{o}noma de Barcelona, 08193 
Bellaterra (Barcelona) Spain} 
\author{Alexander Iomin} 
\affiliation{Department of Physics, Technion, Haifa, 32000, Israel}
\author{Daniel Campos} 
\affiliation{Grup de F{\'i}sica Estad\'{i}stica.  
Departament de F{\'i}sica. Facultat de Ci{\`e}ncies. 
Edifici Cc. Universitat Aut\`{o}noma de Barcelona, 08193 
Bellaterra (Barcelona) Spain} 
\author{Werner Horsthemke}
\affiliation{Department of Chemistry, 
Southern Methodist University,
Dallas, Texas 75275-0314, USA}

\date{\today}

\begin{abstract}
Combs are a simple caricature of various types of natural 
branched structures, which
belong to the category of loopless graphs and consist of
a backbone and branches. 
We study continuous time random walks on combs and
present a generic method to obtain
their transport properties. 
The random walk along 
the branches may be biased, and 
we account for the effect 
of the branches by renormalizing
the waiting time probability distribution function  
for the motion along the backbone. 
We analyze the overall diffusion properties 
along the backbone and find normal diffusion, 
anomalous diffusion, and stochastic localization 
(diffusion failure), respectively,
depending on the characteristics of the 
continuous time random walk along the branches. 
\end{abstract}

\maketitle

\section{Introduction}

Random walks often provide the underlying 
mesoscopic mechanism for transport phenomena
in physics, chemistry and biology \cite{MoSh84,MeKl00,KlSo11}.
A wide class of random walks give rise to normal diffusion,
where the mean-square displacement (MSD), 
$\langle (\Delta r)^{2}(t)\rangle$,
grows linearly with time $t$ for long times.
In many important applications, however, the MSD behaves
like $\langle (\Delta r)^{2}(t)\rangle\propto t^{\gamma}$,
with $\gamma\neq 1$,
and the diffusion is anomalous \cite{MoSh84,MeKl00}.
Anomalous diffusion can be modelled by various classes of random
walks \cite{MeKl04}. We focus on the important class of
continuous time random walks (CTRWs) \cite{MoSh84,MeKl00}.
A specific feature of a CTRW is that 
a walker waits for a random time
$\tau$ between any two successive jumps. 
These waiting times are random independent
variables with a probability distribution 
function (PDF) $\phi(\tau)$, and the tail of the PDF
determines if the transport is
diffusive ($\gamma=1$) or subdiffusive ($\gamma < 1$). 
Heavy-tailed waiting time PDFs give rise to
subdiffusion.
Realistic models of the
waiting time PDF have been formulated for transport in
disordered materials with fractal and ramified
architecture, such as porous discrete media \cite{MaBaShBr03}
and 
comb and dendritic polymers \cite{CaBe66,DoRoFr90,Fr05}, 
and for transport in crowded environments \cite{So12}.

A simple caricature of various types 
of natural branched structures that
belong to the category of loopless graphs 
is a comb model (see Fig.~\ref{fig:f1}).
The comb model was introduced to understand anomalous
transport in percolation clusters \cite{WhBa84,WeHa86,ArBa91}.
Now, comb-like models are widely employed 
to describe various experimental
applications. These models have proven useful
to describe the transport 
along spiny dendrites \cite{MeIo13,IoMe13},
percolation clusters with dangling bonds \cite{WeHa86}, 
diffusion of
drugs in the circulatory system \cite{MaRiMc08}, 
energy transfer in comb
polymers \cite{CaBe66,DoRoFr90} and 
dendritic polymers \cite{Fr05}, diffusion
in porous materials \cite{ArKuBa11,StCo84,TaJa12}, 
the  influence of
vegetation architecture on the diffusion of 
insects on plant surfaces
\cite{Ha02}, and many other interdisciplinary applications. 
Random walks on comb structures provide
a geometrical explanation
of anomalous diffusion.

More general combs have been studied recently. 
For example,
a numerical study of the encounter problem of 
two walkers
in branched structures shows that the topological 
heterogeneity of
the structure can play an important role \cite{AgBlCa14}. 
Another example
is the occupation time statistics for random walkers 
on combs
where the branches
can be regarded as independent complex structures, 
namely fractal or other
ramified branches \cite{ReBa13}. Finally, we want to 
mention studies
to understand the diffusion mechanism along 
a variety of branched
systems, where scaling arguments, verified by 
numerical simulations, have
been able to predict how the MSD
grows with time \cite{FoBuCeVu13}.

Diffusion on comb structures has also been 
studied by
macroscopic approaches, 
based on
Fokker-Planck equations \cite{ArBa91},
which have been applied 
to describe diffusive properties 
in discrete systems, such as
porous discrete media \cite{MaBaShBr03}, 
infiltration of diffusing particles
from one material into another \cite{KoBa10}, and 
superdiffusion due
to the presence of inhomogeneous
convection flow \cite{BaIo04,IoBa05}.
Other macroscopic descriptions, based 
on renormalizing the waiting
time PDF for jumps along
the backbone
to take into account the 
transport along the branches
\cite{vdb89}, have been found
useful 
to model 
continuous-time-reaction-transport 
processes \cite{CaMe05} and 
human migrations along river networks \cite{CaFoMe06}.

Kahng and Redner provided a mesoscopic, 
probabilistic description of random walks 
on combs, by using the
successive decimation of the discrete-time Master 
equation to obtain a
mesoscopic balance equations for the probability of 
the walker to be at
a given node at a given time \cite{KaRe89}. 
A mesoscopic description 
is necessary for
 an accurate description of the transport 
properties, such as
the diffusion coefficient or the mean visiting 
time in a branch, in terms
of the microscopic parameters that
characterize the random walk 
process.

Here we obtain transport 
quantities within the framework of the CTRW
formalism. We assume that the
motion along the backbone
and the branches is non-Markovian and 
that the motion
along the branches can be non-isotropic.
We reduce the dynamic effect of the branches 
to a
waiting time PDF
for the motion along the backbone by using the 
decimation method of Kahng and Redner.
The time spent by the walker between its entry 
into the branches and its return to
the backbone for the first time is
treated as a contribution to the 
effective waiting time at the node
where the branch crosses the backbone.

The paper is organized as follows. 
In Sec.~\ref{sec:mdesc} we formulate
the mesoscopic description of the random walk
on the comb and reduce walker's motion 
to an effective motion
along the backbone only with
a renormalized waiting time PDF for
the backbone nodes. Sec.~\ref{sec:stat} deals
with the MSD of the effective backbone motion,
derives the effective diffusion coefficient, 
and establishes the conditions
for normal diffusion, anomalous
diffusion, and stochastic localization
(diffusion failure) \cite{DeHo00}
in terms of
the number of branch nodes and the degree
of bias of the motion along the branches.
We provide details of the numerical calculations
in Sec.~\ref{sec:numer} and summarize our
results and discuss their 
implications in Sec.~\ref{sec:concl}.

\section{Mesoscopic description}\label{sec:mdesc}

\begin{figure}[htbp]
\includegraphics[width=0.7\hsize]{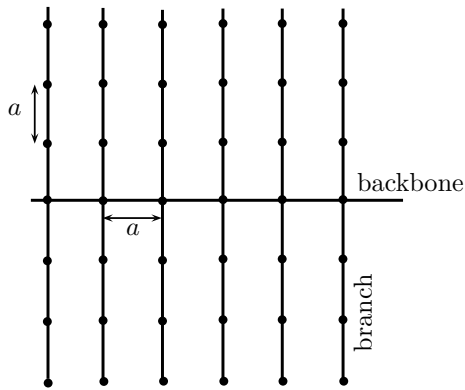}
\caption{Comb structure consisting of a backbone 
and branches. Each point represents a 
node where the walker may jump or wait a random time}
\label{fig:f1}
\end{figure}

The simplest comb model, shown in Fig. \ref{fig:f1}, 
is formed by a principal axis, called the
backbone, which is a one-dimensional 
lattice with spacing $a$,  and identical branches
that
cross the backbone perpendicularly at each node.
The walker moves through the comb by
performing jumps between
nearest-neighbor nodes along 
the backbone or along the branches.
We assume that
the walker performs isotropic jumps along the backbone, 
but 
the
jumps along the branches may be biased,
for example by an external field \cite{WhBa84}.

We derive the balance equation for the PDF $P(x,t)$ 
of finding
the walker at node $x$ on the backbone at time $t$. 
When the walker arrives at a node, 
it waits a random time $\tau$
before performing a new jump to the nearest node. 
We assume that the comb is homogeneous,
and the waiting time PDF at any given node
is given by $\phi_{0}(\tau)$.
When the walker enters a branch, it spends some time
moving inside the branch
before returning to the backbone. 
This sojourn time can be used to
determine an effective waiting time PDF
$\phi(\tau)$ for the walker's motion along
the backbone.
In other words,  
the motion of the walker on the comb
can be reduced
to the effective
motion along a one-dimensional lattice, 
corresponding to the backbone only.
This motion is non-Markovian and can be
described mesoscopically by 
the Generalized Master Equation
(GME)
\begin{multline}
\frac{\partial P}{\partial t}=\int_{0}^{t}K(t-t')dt'\times\\
\left[\int_{-\infty}^{\infty} P(x-x',t')\Phi(x')dx'-P(x,t')\right],
\label{eq:me}
\end{multline}
where $K(t)$ is the memory kernel related to the waiting time PDF
via its Laplace transform, 
$K(s)=s{\phi}(s)/[1-{\phi}(s)]$,
where $s$ is the Laplace variable. 
The dispersal kernel $\Phi(x)$
represents the probability for the walker 
of performing a jump of
length $x$. If the walker moves 
isotropically between
nearest neighbors in a one-dimensional
lattice of spacing $a$, 
the dispersal kernel reads 
$\Phi(x)=\delta(x-a)/2+\delta(x+a)/2$.
We assume that the walker is initially
located at $x=0$, i.e., $P(x,0)=\delta_{x,0}$
with $x=ia$ and $i=0,\pm 1,\pm 2, \dotsc$, where 
$\delta_{x,0}$ is the Kronecker delta.
Then the Laplace transform of
the GME for $x\neq 0$
reads
\begin{equation}
P(x,s)=\frac{\phi(s)}{2}\left[P(x-a,s)+P(x+a,s)\right].
\label{eq:me2}
\end{equation}

\begin{figure}[htbp]
\includegraphics[width=0.5\hsize]{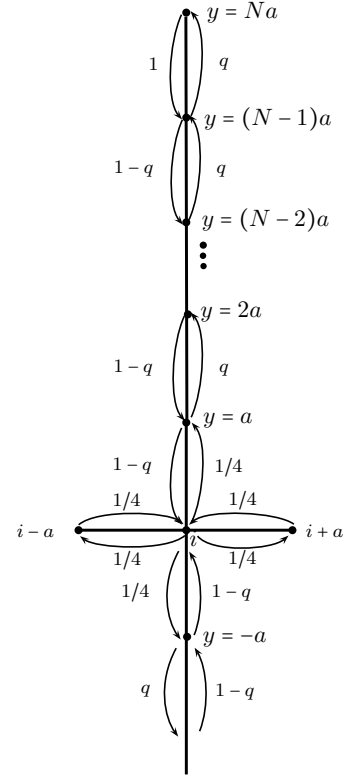}
\caption{Schematic representation of the possible jumps 
of a walker with the corresponding probabilities.}
\label{fig:f2}
\end{figure}

The mesoscopic balance equation for the walker 
on the comb
being at node $x=ia$
of the backbone is
\begin{multline}\label{eq:me3}
P(x,s)=\frac{\phi_{0}(s)}{4}\left[P(x-a,s)+P(x+a,s)\right] \\
+(1-q)\phi_{0}(s)\left[P(y=a,s)+P(y=-a,s)\right].
\end{multline}
Here $P(x,s)$, $P(x-a,s)$, and $P(x+a,s)$
is short-hand for 
$P(x,y=0,s)$, $P(x-a,y=0,s)$, and $P(x+a,y=0,s)$,
and $P(y=a,s)$ and $P(y=-a,s)$ stands for
$P(x,y=a,s)$ and $P(x,y=-a,s)$.
The term $\phi_{0}(s)\left[P(x-a,s)+P(x+a,s)\right]/4$ 
corresponds to
the contribution of 
the walker arriving at node $x=ia$ from the left
or from the right with probability 1/4 
after waiting a random time $\tau$ with PDF
$\phi_{0}(\tau)$ at nodes $x+a$ or $x-a$. 
As shown in Fig.~\ref{fig:f2},
the walker located at
the $i$th node of
the backbone may jump to the right, left, up 
or down with probability
1/4. We assume that the walker moves forward 
(away from the backbone)
along the
branches with probability $q$ and back to the
backbone with probability $1-q$.
The term 
\begin{equation}
(1-q)\phi_{0}(s)\left[P(y=a,s)+P(y=-a,s)\right]
\end{equation}
in (\ref{eq:me3}) corresponds the contribution of 
the walker arriving
at the backbone node $x$ 
from the first node of the upper or 
lower branch after waiting 
there a random time
$\tau$ with PDF
$\phi_{0}(\tau)$.

Consider the motion along the upper branches. 
The lower branch
dynamics is the same due to the symmetry of the comb. 
The mesoscopic balance
equation for the first node of the upper branches reads
\begin{equation}
\label{eq:me4}
P(y=a,s)=\frac{\phi_{0}(s)}{4}P(x,s)+\phi_{0}(s)(1-q)P(y=2a,s).
\end{equation}
The first term $\phi_{0}(s)P(x,s)/4$ corresponds to
the 
contribution of the walker
arriving from the backbone,
while $\phi_{0}(s)(1-q)P(y=2a,s)$ is
the contribution of the walker jumping 
from the upper node $y=2a$
to $y=a$ with probability $1-q$ 
after waiting  a random time
$\tau$ with PDF
$\phi_{0}(\tau)$.
Analogously, we have for the lower branches
\begin{equation}
\label{eq:me4b}
P(y=-a,s)=\frac{\phi_{0}(s)}{4}P(x,s)+\phi_{0}(s)(1-q)P(y=-2a,s).
\end{equation}
Generalizing (\ref{eq:me4}) to any node of the branches 
located between  $2a\leq y\leq(N-2)a$, 
we obtain the balance equation
for the upper branches
\begin{equation}
\label{eq:me5}
P(y,s)=\phi_{0}(s)\left[qP(y-a,s)+(1-q)P(y+a,s)\right].
\end{equation}

To determine the Laplace transform $\phi(s)$
of the effective backbone node waiting
time PDF, we need to determine 
$P(y=a,s)$ and $P(y=-a,s)$
in \eqref{eq:me3}
in terms of $P(x,t)$, so that
\eqref{eq:me3} can be cast in the form
of \eqref{eq:me2}. Given \eqref{eq:me4}
and \eqref{eq:me4b}, this goal can be achieved
if $P(y=2a,s)$ and $P(y=-2a,s)$
can be related 
to $P(y=a,s)$ and $P(y=-a,s)$.
We proceed as follows.
The solution of (\ref{eq:me5}) reads
\begin{equation}
\label{eq:solme5}
P(y,s)=A_{1}\lambda_{+}^{y/a}+A_{2}\lambda_{-}^{y/a},
\end{equation}
where
\begin{equation}
\lambda_{\pm}=\frac{1\pm\sqrt{1-4q(1-q)\phi_{0}^{2}(s)}}
{2(1-q)\phi_{0}(s)}.
\end{equation}
To find expressions for
the quantities $A_{1}$ and $A_{2}$,
whose dependence on $x$ and $s$ is not displayed,
we apply 
(\ref{eq:solme5}) to the node $y=2a$:
\begin{equation}\label{eq:sol2me5}
P(y=2a,s)=A_{1}\lambda_{+}^{2}+A_{2}\lambda_{-}^{2}.
\end{equation}
On the other hand, setting $y=2a$ in (\ref{eq:me5}), 
we find
\begin{multline}
P(y=2a,s)=\phi_{0}(s)\left[qP(y=a,s)\right.\\
+\left.(1-q)\phi_{0}(s)P(y=3a,s)\right],
\end{multline}
or
\begin{multline}
P(y=2a,s)-\phi_{0}(s)qP(y=a,s)=\\
\phi_{0}(s)(1-q)\phi_{0}(s)P(y=3a,s).
\end{multline}
Setting $y=3a$
in (\ref{eq:solme5}) we obtain
\begin{multline}
\label{eq:sol3me5}
P(y=2a,s)-q\phi_{0}(s)P(y=a,s)\\
=\phi_{0}(s)(1-q)\left[A_{1}\lambda_{+}^{3}+A_{2}\lambda_{-}^{3}\right].
\end{multline}
Solving the system of equations (\ref{eq:sol2me5}) and (\ref{eq:sol3me5})
for the quantities $A_{1}$ and $A_{2}$, we obtain
\begin{multline}
\label{eq:a1}
A_{1}=\frac{P(y=2a,s)-q\phi_{0}(s)P(y=a,s)}
{\lambda_{+}^{2}\left(\lambda_{+}-\lambda_{-}\right)\phi_{0}(s)(1-q)}\\
-\frac{\lambda_{-}P(y=2a,s)}{\lambda_{+}^{2}\left(\lambda_{+}-\lambda_{-}\right)},
\end{multline}
\begin{multline}
\label{eq:a2}
A_{2}=\frac{-P(y=2a,s)+q\phi_{0}(s)P(y=a,s)}
{\lambda_{-}^{2}\left(\lambda_{+}-\lambda_{-}\right)\phi_{0}(s)(1-q)}\\
+\frac{\lambda_{+}P(y=2a,s)}{\lambda_{-}^{2}\left(\lambda_{+}-\lambda_{-}\right)}.
\end{multline}

A special situation occurs at the end of the branches,
where we have
to impose reflecting boundary conditions, i.e.,
\begin{equation}
\label{eq:me6}
P(y=Na,s)=q\phi_{0}(s)P(y=(N-1)a,s).
\end{equation}
The node at $y=(N-1)a$ also needs a special balance equation
(see
Fig.~\ref{fig:f2}),
\begin{multline}
\label{{eq:me7}}
P(y=(N-1)a,s)=q\phi_{0}(s)P(y=(N-2)a,s)\\
 +\phi_{0}(s)P(y=Na,s).
\end{multline}

Substituting $y=(N-2)a$ into (\ref{eq:me5}) 
and considering (\ref{eq:me6}),
we can write
\begin{equation}
\label{eq:p2p3}
P(y=(N-2)a,s)=h(\phi_{0}(s))P(y=(N-3)a,s),
\end{equation}
where
\begin{equation}
\label{eq:h}
h(\phi_{0}(s))=\frac{q\phi_{0}(s)\left[1-q\phi_{0}^{2}(s)\right]}
{1+q(q-2)\phi_{0}^{2}(s)}.
\end{equation}
Substituting 
the solutions from
(\ref{eq:solme5}), (\ref{eq:a1}), and (\ref{eq:a2})
into (\ref{eq:p2p3}),
we find
\begin{equation}
\label{eq:p2p1}
P(y=2a,s)=G(q,\phi_{0}(s))P(y=a,s),
\end{equation}
where
\begin{align}
\label{eq:gh}
G(q,\phi_{0}(s))&=\frac{2q\phi_{0}(s)}{1+
\dfrac{1+H(q,\phi_{0}(s))}{1-H(q,\phi_{0}(s))}
\sqrt{1-4q(1-q)\phi_{0}^{2}(s)}},\\
H(q,\phi_{0}(s))&=\left(\frac{\lambda_{-}}{\lambda_{+}}\right)^{N-5}
\frac{\lambda_{-}-h(\phi_{0}(s))}{\lambda_{+}-h(\phi_{0}(s))}.
\end{align}
For the lower branch we obtain in
a similar manner,
\begin{equation}
\label{eq:pm2m1}
P(y=-2a,s)=G(q,\phi_{0}(s))P(y=-a,s).
\end{equation}
We have achieved our goal of
expressing
$P(y=2a,s)$ and $P(y=-2a,s)$
in terms of $P(y=a,s)$ and $P(y=-a,s)$.
Substituting  (\ref{eq:p2p1}) and (\ref{eq:pm2m1}) 
into (\ref{eq:me4})
and (\ref{eq:me4b}) and 
using the resulting expressions in (\ref{eq:me3}),
we obtain an equation of the
form (\ref{eq:me2}) with
\begin{equation}
\label{eq:wt}
\phi(s)=\frac{\phi_{0}(s)}{2-
\dfrac{(1-q)\phi_{0}^{2}(s)}{1-(1-q)\phi_{0}(s)G(q,\phi_{0}(s))}}.
\end{equation}
The Laplace inversion of (\ref{eq:wt}) yields 
$\phi (\tau)$, which incorporates the dynamics 
along the branches and can be understood as the
effective waiting time PDF for a walker
moving along the backbone only.

\section{Statistical properties}\label{sec:stat}
\subsection{$N$ finite}

\begin{figure}[htbp]
\includegraphics[width=1.\hsize]{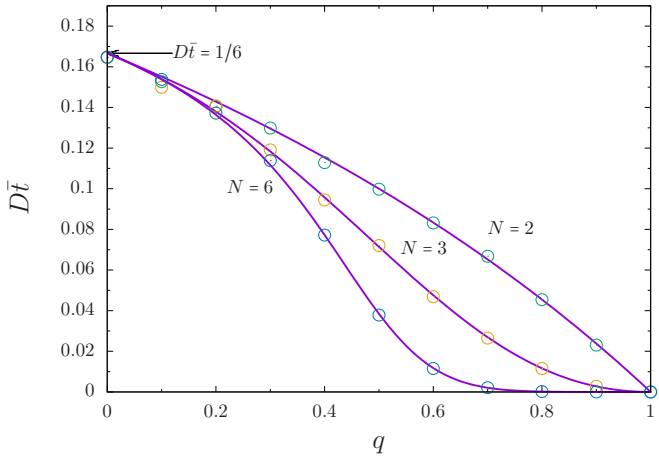}
\caption{Plot of the diffusion coefficient through the comb
for $N=2$, $N=3$, and $N=6$ versus $q$.
Solid curves correspond to exact analytical
results given by (\ref{eq:d}).
Numerical simulations results are depicted with symbols.}
\label{fig:f4}
\end{figure}

If the local waiting time PDF $\phi_{0}(\tau)$ has 
finite moments, its Laplace transform  
reads \cite{MeKl00}, $\phi_{0}(s)\simeq1-s\bar{t}$, 
in the large time limit $s\rightarrow0$,
where $\bar{t}$ is the local mean waiting 
time at each node. Taking the limit
$s\rightarrow0$ in (\ref{eq:wt}), 
we obtain the waiting time PDF 
for the effective backbone
dynamics,
\begin{equation}\label{eq:phi}
\phi(s)\simeq(1+s\left\langle t\right\rangle )^{-1}. 
\end{equation}
The mean waiting time $\left\langle t\right\rangle$
is given by
\begin{equation}
\label{eq:mwt}
\left\langle t\right\rangle =
\frac{\bar{t}}{2q-1}\left[2(1-q)^{1-N}q^{N}+4q-3\right].
\end{equation}
\begin{figure}[htbp]
\includegraphics[width=0.9\hsize]{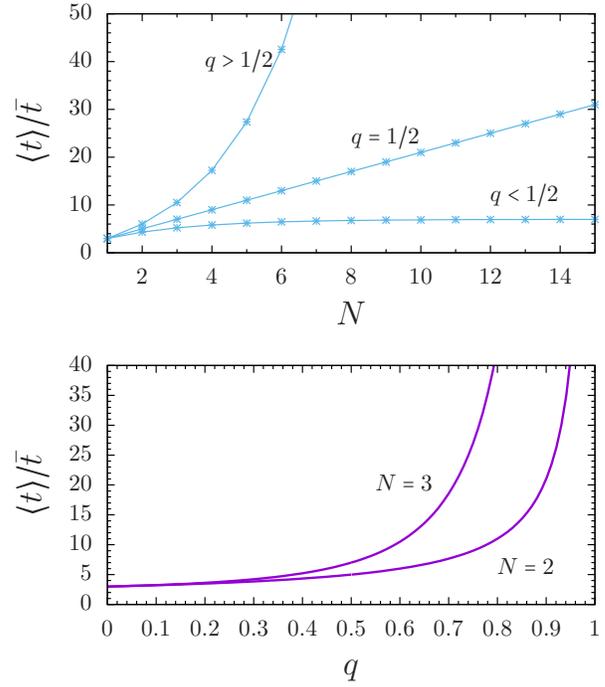}
\caption{Dimensionless
mean waiting time of the
effective backbone dynamics.}
\label{fig:f3}
\end{figure}
In Fig.~\ref{fig:f3},
we plot the effective mean waiting at a node
of the backbone dynamics
versus $N$ and $q$. 
It shows that the 
mean waiting time $\left\langle t\right\rangle$
is a monotonically increasing function of both $q$ and $N$.
If the random walk inside the branches is isotropic,
$q=1/2$, 
one obtains by the L'Hopital's rule from (\ref{eq:mwt})
\begin{equation}
\lim_{q\rightarrow1/2}\left\langle t\right\rangle 
=\left(1+2N\right)\bar{t}.
\end{equation}
To determine the diffusion coefficient $D$ 
for diffusion through the comb,
we first calculate the MSD.
Performing the Fourier-Laplace transform 
on (\ref{eq:me}), we obtain
\begin{equation}\label{eq:me-FL}
P(k,s)=\frac{1-\phi(s)}{s[1-\Phi(k)\phi(s)]}.
\end{equation}
The MSD in Laplace 
space reads (see, e.g., \cite{MeKl00})
\begin{equation}
\label{eq:msd}
\left\langle x^{2}(s)\right\rangle = 
-\lim_{k\rightarrow0}\frac{d^{2}P(k,s)}{dk^{2}}.
\end{equation}
As mentioned in Sec.~\ref{sec:mdesc},
we assume that the motion on the
backbone is unbiased and that the walker
only jumps to nearest neighbors.
This implies that the kernel $\Phi(x)$ 
is given $\Phi(x)=\delta(x-a)/2+\delta(x+a)/2$, 
and we obtain from
(\ref{eq:msd}),
\begin{equation}
\label{eq:msd2}
\left\langle x^{2}(s)\right\rangle 
=\frac{a^{2}}{s\left[\phi(s)^{-1}-1\right]}\, 
\end{equation}
in the large time limit. 
If
the waiting time PDF
$\phi(t)$ possesses a finite first moment,
(\ref{eq:phi}) implies that the
MSD along the backbone corresponds 
to normal diffusion 
$\left\langle x^{2}(t)\right\rangle =2Dt$.
The diffusion coefficient is given by
\begin{equation}\label{eq:d}
 D=\frac{a^{2}}{2\left\langle t\right\rangle} 
 =\frac{a^{2}}{2\bar{t}}\frac{2q-1}{2(1-q)^{1-N}q^{N}+4q-3}.
\end{equation}
In Fig.~\ref{fig:f4}, we compare the 
results provided by (\ref{eq:d})
with numerical simulations. 

As Fig.~\ref{fig:f3} demonstrates,
$\left\langle t\right\rangle$
increases monotonically with $N$ 
for $q<1/2$ and saturates at
$(4q-3)/(2q-1)$
for $N\to\infty$. Consequently, the
mean waiting time $\left\langle t\right\rangle$
is finite for $N\to\infty$; 
the overall 
diffusion along the backbone
is normal. 
However, for $q\geq1/2$, the mean waiting time
$\left\langle t\right\rangle$ 
increases without bound as $N$ increases,
and anomalous transport
is expected for $N\to\infty$.
\begin{figure}[htbp]
\includegraphics[width=1.\hsize]{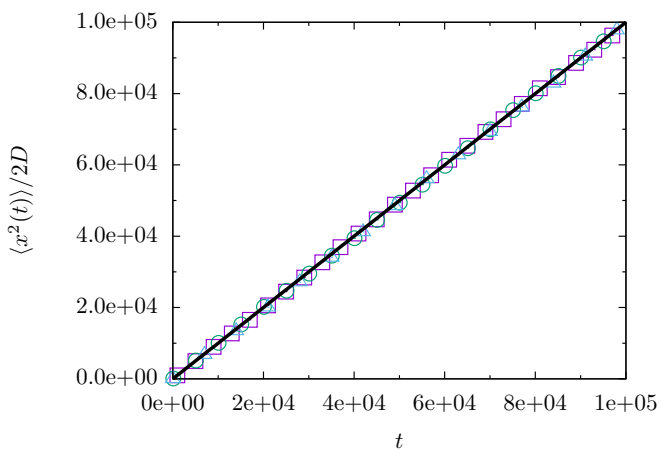}
\caption{MSD over $2D$ for $N$ fixed and three different
values of $q$: 0.1, 0.25, 05. 0.75}
\label{fig:f5}
\end{figure}
In Fig.~\ref{fig:f5}, we plot the MSD scaled 
by the diffusion coefficient. 
It illustrates the result given by (\ref{eq:d}) for the MSD.
The transport is diffusive for finite $N$, regardless $q$ and
the specific form of $\phi_{0}(\tau)$, as 
long as it has finite moments.

We consider now a an effective
waiting time PDF with
the large-time limit 
$\phi_{0}(\tau)\sim \tau^{-1-\gamma}$, 
with Laplace transform
$\phi_{0}(s)\simeq 1-(s\tau_{0})^{\gamma}$
and $0<\gamma<1$, which
does not possess finite moments.  
Here $\tau_{0}$ is a parameter with units of time.
In this case, the waiting time PDF 
for the backbone dynamics is obtained by
simply replacing $s\bar{t}$ with $(s\tau_{0})^{\gamma}$, 
i.e, $\phi(s)\simeq[1+\left(s\tau_{0}\right)^{\gamma}
\left\langle t\right\rangle /\tau_{0}]^{-1}$. 
Substituting this result into (\ref{eq:msd2}),
we find
\begin{equation}
\left\langle x^{2}(t)\right\rangle 
=\frac{a^{2}\tau_{0}}{\left\langle t\right\rangle }
\frac{(t/\tau_{0})^{\gamma}}{\Gamma(1+\gamma)},
\end{equation}
for large $t$, where $\left\langle t\right\rangle $ 
is given by (\ref{eq:mwt}), with $\tau_{0}$ instead
of 
$\bar{t}$.
If the waiting time PDF $\phi_{0}(\tau)$
at each node of the comb has a
power-law tail, 
then the overall transport along 
the backbone is anomalous.

\subsection{$N\to\infty$}

If the number of nodes of the branches 
goes to infinity,
the mean time spent by the walker visiting 
a branch increases monotonically,
see \eqref{eq:mwt}. 
However, this does not always
results in anomalous transport 
along the overall structure as
we show below.

\begin{figure}[htbp]
\includegraphics[width=1.\hsize]{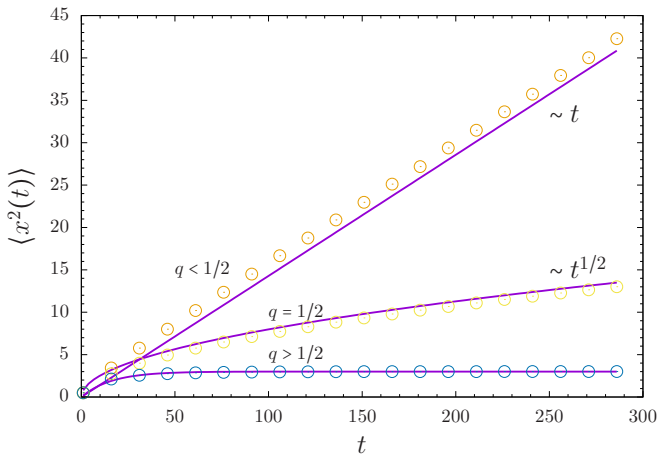}
\caption{MSD for three values of $q$, 
displaying three different behaviors. 
Solid curves correspond to the results 
given by (\ref{eq:msdi}). 
Symbols are the results of numerical 
simulations with $N=10^3$.}
\label{fig:f6}
\end{figure}

For $N\to\infty$, 
the quotient $(\lambda_{-}/\lambda_{+})^{N}\to 0$
and also $H\to 0$. We obtain from 
(\ref{eq:gh}),
\begin{equation}
G(q,\phi_{0}(s))=\frac{2q\phi_{0}(s)}
{1+\sqrt{1-4q(1-q)\phi_{0}^{2}(s)}}\equiv
\frac{2q\phi_{0}(s)}{1+g(q)},
\end{equation}
where we define 
$g(q)\equiv\sqrt{1-4q(1-q)\phi_{0}^{2}(s)}$ 
for  convenience.
Equation (\ref{eq:wt}) reduces to
\begin{equation}
\label{eq:fis}
\phi(s)=\frac{\phi_{0}(s)
\left[1+g(q)-2q(1-q)\phi_{0}^{2}(s)\right]}
{2-(1+3q-4q^2)\phi_{0}^{2}(s)+
[2-(1-q)\phi_{0}^{2}(s)]g(q)}.
\end{equation}
We take the limit $s \to 0$ and
consider first the case where
$\phi_{0}(\tau)$ has finite moments. 
Then $\phi_{0}(s)\simeq 1-s\bar{t}$,
as $s\to 0$. The square root $g(q)$ in
(\ref{eq:fis}) reads
\begin{equation}
g(q)\simeq
\begin{cases}
1-2q-\dfrac{4q(1-q)}{2q-1}s\bar{t}, & q<1/2,\\[2ex]
\sqrt{2s\bar{t}}-\dfrac{\sqrt{2}}{4}(s\bar{t})^{3/2}, & q=1/2,\\[2ex]
-1+2q+\dfrac{4q(1-q)}{2q-1}s\bar{t}, & q>1/2,
\end{cases}
\end{equation}
and (\ref{eq:fis}) implies that the waiting time PDF 
is given by
\begin{equation}
\label{eq:fis2}
\phi(s)\simeq
\begin{cases}
\left(1+\dfrac{4q-3}{2q-1}s\bar{t}\right)^{-1}, & q<1/2,\\[2ex]
\left(1+\sqrt{2s\bar{t}}\right)^{-1}, & q=1/2,\\[2ex]
\left(\dfrac{3q-1}{q}+\dfrac{4q^{2}-3q+1}{(2q-1)q}s\bar{t}\right)^{-1}, & q>1/2.
\end{cases}
\end{equation}
Substituting (\ref{eq:fis2}) into (\ref{eq:msd2}),
we find for large $t$,
\begin{equation}
\label{eq:msdi}
\left\langle x^{2}(t)\right\rangle =
\begin{cases}
a^{2}\dfrac{2q-1}{4q-3}\dfrac{t}{\bar{t}}, & q<1/2,\\[2ex]
a^{2}\sqrt{\dfrac{2t}{\pi\bar{t}}}, & q=1/2,\\[2ex]
a^{2}\dfrac{q}{2q-1}\left(1-e^{-\alpha t}\right),& q>1/2,
\end{cases}
\end{equation}
where the rate of saturation is 
\begin{equation}
\alpha = \frac{(2q-1)^{2}}{(4q^{2}-3q+1)\bar{t}}.
\end{equation}
In Fig~\ref{fig:f6} we compare these
results with numerical simulations
for $N=10^{3}$.
For $q=1/2$, we obtain the 
well known result of subdiffusive transport
with the MSD $\sim \sqrt{t}$.
However, for $q\neq 1/2$, the side branches experience
advection, and the transport is remarkably different. 
Namely, for $q>1/2$ the advection is 
away from the backbone 
along
the branches, $y\rightarrow\pm\infty$.
The walker is effectively trapped 
inside the branches, 
and 
stochastic localization (diffusion failure)
occurs, 
$\left\langle x^{2}(\infty)\right\rangle
< \infty$,
\cite{DeHo00}.
For $q<1/2$, the advection
is towards the backbone.
It enhances the backbone dynamics 
and normal diffusion takes place.

Consider now the case where the 
local waiting time PDF is $\phi_{0}(\tau)
\sim \tau^{-1-\gamma}$, i.e.,
$\phi_{0}(s)\simeq1-(s\tau_{0})^{\gamma}$ 
with $0<\gamma<1$, as $s\to 0$.
The MSD in this case  
can be obtained straightforwardly 
by replacing $s\bar{t}$ with $(s\tau_{0})^{\gamma}$
in (\ref{eq:fis2}). For large times it reads
\begin{equation}
\label{eq:msd-gamma}
\left\langle x^{2}(t)\right\rangle =
\begin{cases}
\dfrac{a^{2}}{\Gamma(1+\gamma)}\dfrac{2q-1}{4q-3}\left(\dfrac{t}{\tau_{0}}\right)^{\gamma}, & q<1/2,\\[2ex]
\dfrac{a^{2}}{\sqrt{2}\Gamma(1+\gamma/2)}\left(\dfrac{t}{\tau_{0}}\right)^{\gamma/2}, & q=1/2,\\[2ex]
\dfrac{a^{2}q(2q-1)}{\tau_{0}^{\gamma}(4q^{2}-3q+1)}\mu(t/\tau_{0}), & q>1/2,
\end{cases}
\end{equation}
where 
\begin{equation}
\mu(t/\tau_{0})=(t/\tau_{0})^{\gamma}
E_{\gamma,\gamma+1}\left[-\left(\frac{t}{\tau_{0}}\right)^{\gamma}
\frac{(2q-1)^{2}}{4q^{2}-3q+1}\right]
\end{equation}
is expressed in terms
of the the generalized 
Mittag-Leffler function $E_{\alpha,\beta}(z)$.
We use the following property of integration 
of the Mittag-Leffler function \cite{Po99},
\begin{equation}
\int_0^tE_{\alpha,\beta}\left(bz^{\alpha}\right)z^{\beta-1}dz=t^{\beta}
E_{\alpha,\beta+1}\left(bt^{\alpha}\right).
\end{equation}
Subdiffusion in the branches results in 
backbone subdiffusion for $q\leq 1/2$. 
For advection away from the backbone,
$q>1/2$, we again
find stochastic localization.
For $t/\tau_{0}\gg 1$,
$E_{\alpha,\beta}(-at^{\alpha})
\sim t^{-\alpha}/\Gamma(\beta-\alpha)$ 
\cite{Ba53}, and
consequently
$\mu(t/\tau_{0})$ approaches
a finite value as $t\to\infty$.

\section{Conclusion}\label{sec:concl}

We have developed a mesoscopic equation 
for a random walk on a regular comb structure 
given by (\ref{eq:me2}) and (\ref{eq:wt}). 
The random walk along the branches consists 
of, possibly biased, jumps to the 
nearest node, while waiting 
at each node for a random time $\tau$ 
distributed according to the PDF $\phi_0(\tau)$ 
before proceeding with the next jump. 
The overall dynamics along the branches 
has been reduced to an
effective waiting time PDF,
given by (\ref{eq:wt}),
for motion solely 
along the backbone. We have obtained  
statistical properties, such as the 
effective mean 
waiting time, $\langle t\rangle$ for
the backbone nodes,
and the diffusion coefficient, $D$, of 
the overall structure for the case where 
the number of nodes $N$ of the branches 
is finite or infinite. 
If $N$ is finite and $\phi_0(\tau)$ 
has finite moments, both $\langle t\rangle$ and
$D$ are derived exactly in terms of the bias 
probability $q$, the number
of nodes $N$ on the branch,
and the mean waiting time 
probability at each node. 
In this case the
transport is always
normal diffusion. 
If $\phi_0(\tau)\sim \tau^{-1-\gamma}$ 
for large time, it does not
posses finite moments and the
MSD of the random walker
behaves like $t^{\gamma}$.
If $N$ is infinite,
the value of $q$ is decisive. 
If $\phi_0(\tau)$ has finite moments, 
the diffusion 
regime is normal if $q<1/2$, while the 
MSD behaves like $t^{1/2}$ for $q=1/2$. 
If $q>1/2$, 
the MSD approaches a constant finite
value 
for large time, corresponding
to stochastic localization (diffusion
failure). 
If $\phi_0(\tau)\sim \tau^{-1-\gamma}$ 
for large time, the MSD behaves
like $t^{\gamma}$ 
for $q<1/2$ and like $t^{\gamma/2}$ for $q=1/2$ 
Again, stochastic localization
occurs for $q>1/2$. 
In summary, if the bias probability of moving 
away from the backbone is $q>1/2$, then
stochastic localization occurs, 
regardless of the other 
characteristic parameters related to the 
random walk on the branches.

\section*{Acknowledgments}

A.I. would like to thank the Universitat Aut\`onoma
de Barcelona for hospitality and financial support, 
as well as the support by the
Israel Science Foundation (ISF-1028). VM and DC have 
been supported by the Ministerio de Ciencia e 
Innovaci\'on under Grant No. FIS2012-32334. 
VM also thanks the Isaac Newton Institute for 
Mathematical Sciences, Cambridge, for support 
and hospitality during the CGP programme where 
part of this work was undertaken. 


\bibliography{combs}

\end{document}